\documentclass{book}    % fundamental class file is 'book'.
\usepackage{piers}  % use the file 'piers.sty'.
\begin{document}

%=== Input Title here ================================
\title{Franck-Hertz experiment in magnetic field}
\maketitle
%===================================================

%=== List of authors (in order) ========
%-- Author(s) for the first affiliation ---
\author      {Ying Weng}
\affiliation {Xiamen University}
\address     {}% optional
\city        {Xiamen}
\postalcode  {}% optional
\country     {China}
\phone       {345566}    % optional
\fax         {233445}    % optional
\email       {xmuwzh@hotmail.com}  % optional
\misc        { }  % optional
\nomakeauthor
%------------------------------------

\author      {Zi-Hua Weng}
\affiliation {Xiamen University}
\address     {}% optional
\city        {Xiamen}
\postalcode  {}% optional
\country     {China}
\phone       {345566}    % optional
\fax         {233445}    % optional
\email       {xmuwzh@xmu.edu.cn}  % optional
\misc        { }  % optional
\nomakeauthor
%------------------------------------

%---Output of Authors----------------------

\begin{authors}

{\bf Ying Weng}$^{1}$, {\bf Zi-Hua Weng}$^{2}$ \\
\medskip
$^{1}$ College of Chemistry \& Chemical Engineering, \\Xiamen University, Xiamen 361005, China\\
$^{2}$ School of Physics and Mechanical \& Electrical Engineering, \\Xiamen University, Xiamen 361005, China\\

\end{authors}
%--------------------------

%---Content of Paper Abstract-----------------------
\begin{paper}

\begin{piersabstract}
The paper studies the impact of applied magnetic field on the inelastic collisions of electrons with argon atoms. In the electron-argon Franck-Hertz experiment, the influence of applied magnetic field emerges complicated features, and is equivalent to that of the temperature. In case the accelerating electric intensity becomes strong enough, enlarging magnetic flux density will be equivalent to the increasing of oven temperature. When the accelerating electric intensity is very weak and the applied magnetic field occupies a dominant position, enhancing magnetic flux density is identical with the decreasing of oven temperature. And the non-uniform distribution of applied magnetic field has an influence on the inelastic collision as well. The study claims that the influence of magnetic field variation is equivalent to that of temperature variety, and that it leads the electron energy to transfer obviously in the experiment.
\end{piersabstract}

%---Content of Paper Text-----------------------

\psection{Introduction}

Franck-Hertz (F-H) experiment is vital to the modern physics, for it confirmed first the existence of discrete energy levels in atoms. In 1914, J. Franck and G. Hertz  \cite{franck} reported an experiment on the collision of electrons with mercury vapor atoms. The authors observed a stepwise loss of electron energy in the experiment. N. Bohr in 1915 adverted to the F-H experiment and brought forward some related advisements. B. Davis and F. S. Goucher \cite{davis} in 1917 modified F-H tube to validate Bohr's predictions about the discrete energy level of gas atoms. In 1919, J. Franck and G. Hertz rechecked prototype F-H experiment, and claimed to support Bohr's viewpoint. Further, they improved their classic experimental devices the next year to study the high excitation level. Subsequently many research experiments \cite{liu, robson} have been carried out on the different features of F-H experiment with the mercury vapor, argon gas, and neon gas etc.

In a F-H tube filled with argon gas, the inelastic collisions of electrons with argon atoms will alter not only collided particles' velocity and direction, but also their energies and physics states. The result is the argon atoms acquire the energy while the electrons loose their energies. Decreasing the energy will cut down the number of electrons reaching the collector, consequently the collected current drops. Along with the increasing of electric voltage, the collected current will be oscillated periodically. And that this variation situation will be repeated again and again. The current variation is reported to be determined by the emitting voltage, accelerating voltage, retarding voltage, and temperature etc. Besides those main factors, the magnetic field is also an important and even more complicated factor.

By means of the air core solenoid and the permanent magnet, the paper studies the applied magnetic field's impact on the inelastic collisions of electrons with argon atoms. And the authors observed a few contrary influences of applied magnetic flux density on different accelerating electric intensity stages. The results obtained show that the influence of applied magnetic field on collected current emerges complicated features in the experiment.

\psection{Influence of temperature}

In the electron-argon collision experiment, the increasing of oven temperature enlarges the electron kinetic energy and collecting current, but it does not vary the mean-free-path of electrons almost. Therefore the increasing of oven temperature leads testing curves to lift up, while the wave troughs of testing curves do not shift approximately.

\psubsection{F-H tube configuration}

The tetrode F-H tube used in the experiment is a cylindrical tube filled with argon gas in Figure 1. The F-H tube configuration includes an indirectly heated cathode K, the heater F, grid G1, grad G2, and collector C. The grids G1 and G2 both are helix rounded by the nickel filaments. The G1 coated with the gold is located in the tube center along the cylindrical axis. The distance between G1 and K is near 0.3 mm. The grad G2 is arranged outside G1 and is 1 mm from the collector C. The distance between G1 and G2 is shorted to be comparable with the mean-free-path of electron in the argon gas at operating temperature. The result shows an excellent characteristic curve in which as many as 6 wave crests can be seen.

The F-H tube is divided into three parts: the emitting region K-G1, the accelerating and collision region G1-G2, and the retarding region G2-C. Selecting a proper pressure of argon gas, the mean-free-path of electrons will be slightly larger than the distance between K and G1. Consequently the electrons will be accelerated without collided in the emitting region. The electrons emitted from the cathode are accelerated by the voltage applied on the grids G1 and G2. The advantage of F-H tube configuration is that a large region between G1 and G2 will increase greatly the collision possibility among electrons and argon atoms. The most electrons collided inelastically with argon atoms loose a large part of their energies, they can not overcome the retarding electric field to reach the collector. Accordingly it drops the collected current. It is clear that when the critical level of argon atoms are reached, a series of successive wave crests and troughs on the voltage-current curves will be obtained in the experiment.

\psubsection{Temperature effect}

The F-H tube is located in an oven, which wrapped with an aluminum electrical shield. The oven temperature is raised to the operating value, and the heater supplies heat to the cathode. A lower cathode temperature will minimize the disturbance of velocities distribution of thermionic emitted electrons. During the experiment, the temperatures of oven and heater are kept constantly, since the collected current is sensitive to these temperatures.

For small accelerating voltages the collected current characteristics of the F-H tube are similar to that of a tetrode. At certain accelerating voltage, however, the collected current reaches a maximum. Increasing the accelerating voltage further decreases the collected current as the cross section for inelastic collision changes with the electron energy. When almost electrons have suffered the inelastic collision, the collected current traverses a minimum, and then it increases again.

It can easily be seen that there are a series of successive wave crests and troughs on the testing curve. In case emitting voltage and retarding voltage both are kept constantly, increasing the oven temperature lifts up the testing curve. When the oven temperature increases, the mean-free-path of electron varies slightly and the requested accelerating voltage keeps the same, according to the formula of mean-free-path of electron, $\lambda = K_b T / p \sigma $. Here $K_b$ is the Boltzmann constant; $T$ and $p$ are the systematic temperature and pressure respectively; $\sigma$ is the cross section for inelastic collision. Therefore the each wave trough of testing curves does not shift almost, due to the mean-free-path of electron keeps the same approximately. Moreover, the testing curves will lift up, because increasing the oven temperature enlarges the electron kinetic energy and the number of electron reaching the collector in Figure 2. Besides the temperature and electric intensity, the magnetic flux density has an influence on the electron-argon inelastic collision in F-H experiment also.

\begin{figure}[h]
\begin{minipage}[]{0.5\linewidth}
\setcaptionwidth{3.in}
\centering
\includegraphics[]{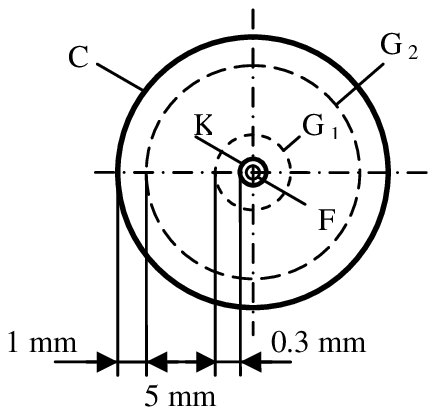}
\caption{Cross-section sketch of a F-H tube with the cylindrical configuration. This tube filled with argon gas includes the heater F, cathode K, grid G1, grid G2, and collector C.}
\label{fig:side:a}
\end{minipage}%
\begin{minipage}[]{0.5\linewidth}
\setcaptionwidth{3.in}
\centering
\includegraphics[]{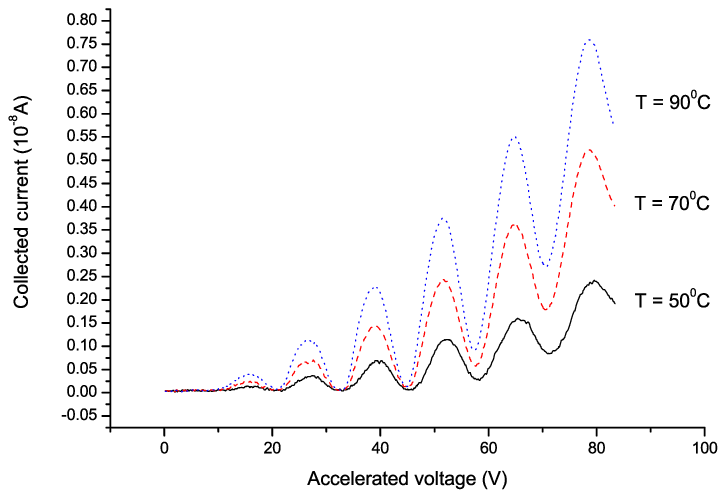}
\caption{Increasing oven temperatures enlarge the electron kinetic energy and lead the voltage-current curves to lift up. $V_f=3.75V, V_{G1K}=0.7V, V_{G2C}=7V$. $T=50^{0}$C, $70^{0}$C, and $90^{0}$C.}
\label{fig:side:b}
\end{minipage}
\end{figure}

\psection{Impact of magnetic field}

In the F-H experiment a uniform magnetic field is applied properly in a direction perpendicular to the accelerating electric intensity. Wrapping the copper enameled wire on the surface of paper cylindrical canister to construct a solenoid, which inducing the uniform magnetic field by applied electric current $I$. Firstly, this solenoid is $9cm$ and longer than the tube length. And it produces a uniform magnetic field to eliminate the disturbance of non-uniform distribution of magnetic field. The non-uniform distribution of applied magnetic field may cause the variation of magnetic potential energy to bother the measure precision. Secondly, the radius of solenoid is larger than that of tube. The F-H tube is located inside the solenoid, and has $5mm$ air interval between them. This solenoid configuration and its paper material both dispel maximally the interference of temperatures. The final results verify the above advisement is effective in fact.

The F-H experiment in uniform magnetic fields is processing in the room temperature. And the influence of applied magnetic field emerges complicated features in the experiment.

\psubsection{Dominant electric intensity}

In the test, the impact of magnetic field variation is equal to that of temperature variety. This indicates that the applied magnetic field results in electrons' energy transferring.

During the F-H experiment, the room temperature and the heating voltage $V_f$ both are kept constantly, when the accelerating voltage $V_{G2K}$ is strong enough and is increasing continually further. There are a series of wave crests and wave troughs on the voltage-current curve, when the emitting voltage $V_{G1K}$ and retarding voltage $V_{G2C}$ both be kept constantly.

When the magnetic flux density $\textbf{B}$ increases, the mean-free-path of electron and the requested accelerating voltage both keep the same. In the experiment, we find that increasing the magnetic flux density lifts up the testing curve, while each wave trough of testing curves does not shift almost in Figure 3. This means that increasing magnetic flux density in short time enlarges the electron kinetic energy in spite of cyclotron emission, consequently the collected current increases. Therefore enlarging magnetic flux density is equivalent to the increasing of oven temperature, when the accelerating electric intensity is strong enough and occupies a dominant position.

\begin{figure}[h]
\begin{minipage}[]{0.5\linewidth}
\setcaptionwidth{3.in}
\centering
\includegraphics[]{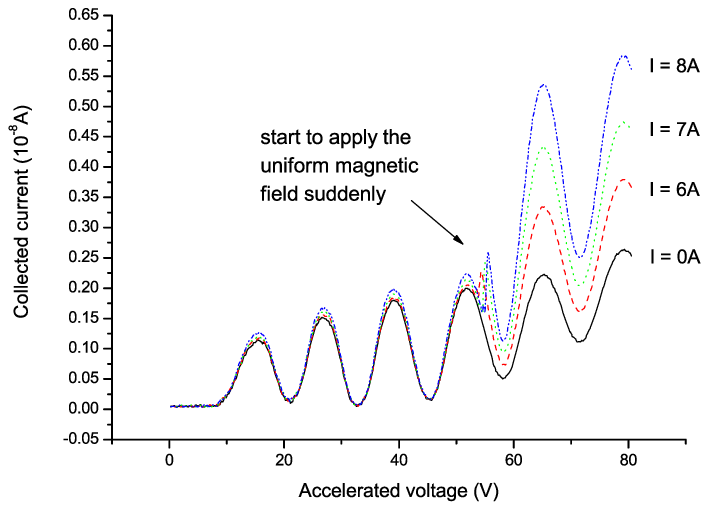}
\caption{When the electric intensity is in a prominent place, increasing the magnetic flux density causes the voltage-current curve to jump up. $V_f=3.78V, V_{G1K}=2.3V, V_{G2C}=7V$. $I=6A$, $7A$, and $8A$, and their $\textbf{B} = 19.5mT$, $23.7mT$, and $26.3mT$ correspondingly.
}
\label{fig:side:a}
\end{minipage}%
\begin{minipage}[]{0.5\linewidth}
\setcaptionwidth{3.in}
\centering
\includegraphics[]{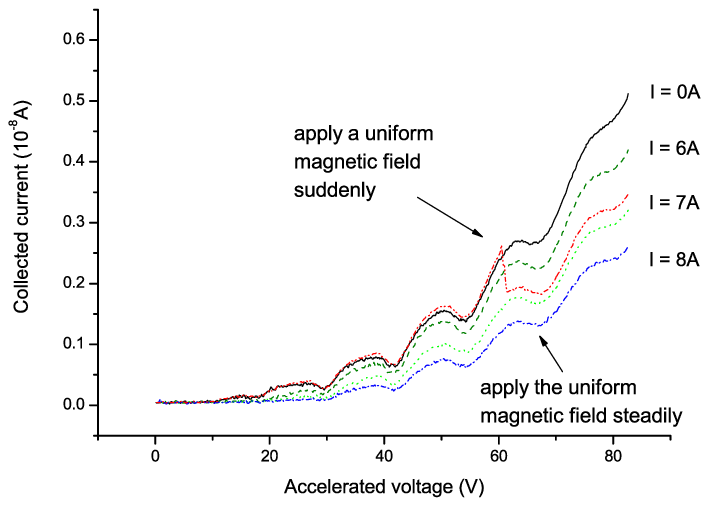}
\caption{When the magnetic flux density occupies a dominant position, enhancing the magnetic flux density lowers the voltage-current curve suddenly or steadily. $V_f=3.9V, V_{G1K}=0.7V, V_{G2C}=1V$. $I=6A$, $7A$, and $8A$, and their $\textbf{B} = 19.5mT$, $23.7mT$, and $26.3mT$ respectively.}
\label{fig:side:b}
\end{minipage}
\end{figure}

\psubsection{Dominant magnetic flux density}

However, enlarging the magnetic flux density in the F-H experiment does not always imply the increasing of collected current, especially in weak accelerated voltages. Introducing a few necessary processing in the test to eliminate the interference coming from oven temperature and non-uniform magnetic field distribution, and protrude directly the effect of uniform magnetic field.

The room temperature, heater temperature, emitting voltage, and retarding voltage are kept constantly in the test. The emitting voltage and accelerating voltage both are weak enough and invariable, meanwhile the magnetic flux density occupies a dominant position. There are a series of wave crests and wave troughs on the testing curve too, when the magnetic flux density is increasing continually further.

When the accelerating electric intensity is very weak and the applied magnetic flux density occupies a dominant position, enhancing magnetic flux density over a short period of time should be identical with the oven temperature decreasing. In the test, we are surprised to find that increasing the magnetic flux density drops down the testing curve in Figure 4. This means that increasing magnetic flux density reduces the number of electron reaching the collector. Analyzing results recovers that the increasing magnetic flux density will enhance the electron kinetic energy and escaping velocity. And then increasing the number of electrons escaping from the tube leads the collected current to decrease. As a result, when the accelerating electric intensity is weak enough and the applied magnetic flux density occupies a dominant position, enlarging magnetic flux density is equivalent to the decreasing of oven temperature. This phenomenon can be explained partly by the secondary electron emission, and partly by the quaternion electromagnetic theory.

\psection{Quaternion electromagnetism}

According to classic electromagnetic theory, the uniform magnetic field has not an influence on the inelastic collisions of electrons with argon atoms, and on the electrons' energy transferring either. However the experiment results contradict this usual assumption. No matter how to apply the uniform magnetic field, either steadily or instantaneously, the collected current will appear to be fluctuated observably.

This means that applying uniform magnetic field leads to electrons' energy fluctuating. But the classic electromagnetic theory does not explain effectively why the phenomena will happen. The research assumes that there may be one new kind of electromagnetic force component, which results in the electrons to be accelerated continually, and then be transferred their energies.

\psubsection{Electromagnetic force}

The quaternion was invented by W. R. Hamilton \cite{hamilton} in 1843, and was first used by J. C. Maxwell in 1861 to represent electromagnetic theory \cite{maxwell}. At present, the algebra of quaternions can be used to describe either electromagnetic field or gravitational field.

The electromagnetic theory \cite{weng} described by quaternions predicts that there exists one new kind of electromagnetic force component, $ q v_0 \textbf{B} $ , along the direction of magnetic field line, besides the Lorentz force etc. Here $ q $ is the electric charge, and $ v_0 $ is the speed of light. This assumed force component will accelerate the electric charge along the direction of magnetic field line, and then vary the electric charge's energy. While Lorentz force causes electrons to turn near the cathode but does not change the electron energy in F-H experiment.

\begin{figure}[h]
\begin{minipage}[]{0.5\linewidth}
\setcaptionwidth{3.in}
\centering
\includegraphics[]{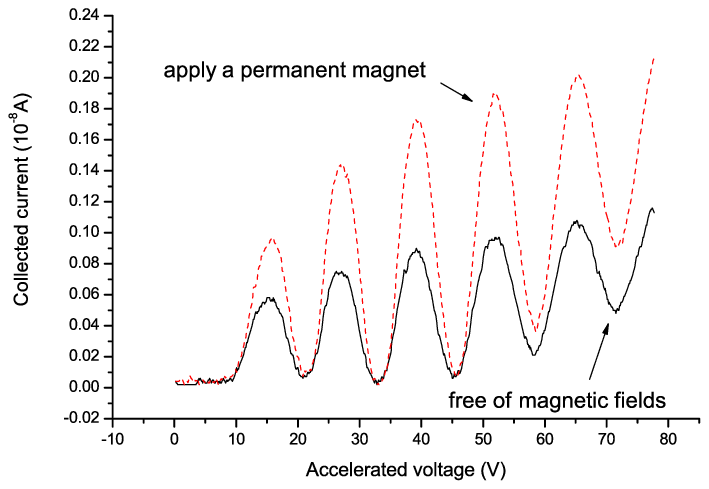}
\caption{Influence of a permanent magnet on the collected current, when the accelerating electric intensity is in a prominent place. $\textbf{B} = 43mT$ . $V_f=3.42V, V_{G1K}=2.3V, V_{G2C}=7.5V$.}
\label{fig:side:a}
\end{minipage}%
\begin{minipage}[]{0.5\linewidth}
\setcaptionwidth{3.in}
\centering
\includegraphics[]{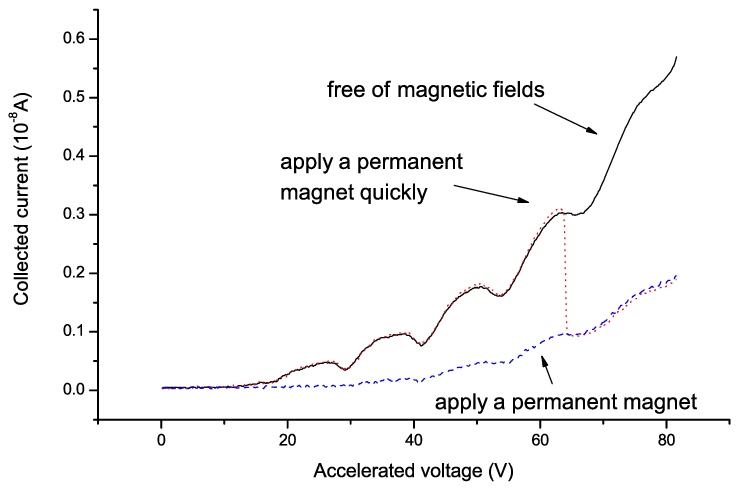}
\caption{Impact of the permanent magnet on the collected current, when the magnetic flux density occupies a dominant position. $\textbf{B} = 43mT$ . $V_f=3.95V, V_{G1K}=0.6V, V_{G2C}=1V$.}
\label{fig:side:b}
\end{minipage}
\end{figure}

\psubsection{Electron escaping}

The influence of applied magnetic field in F-H experiment emerges complicated features. In the test, the impact of magnetic field variation is equal to that of temperature variety. This indicates that the applied magnetic field results in electrons' energy transferring. And it also states that enlarging magnetic flux density leads electrons' energy to increase, when the accelerating electric intensity is strong enough and occupies a dominant position.

In case the accelerating electric intensity occupies a dominant position, the medial electric force is much bigger than the assumed force. The assumed force will accelerate electrons along the magnetic field line, and lift up the electrons' energy by the interparticle collision. Accordingly, it increases electrons' number reaching the collector, and enlarges the collected current in Figure 5. In the same time, the assumed force will cause tiny part of electrons escaping from the F-H tube along the magnetic field lines. Due to the electrons speed from cathode to collector are high enough, the accelerating time is quite short, and then the number of escaping electrons is very small.

Further we lower the accelerated electric intensity while increase the magnetic flux density, it will drop the radius component of electron's speed and increase number of escaping electrons. When the accelerating electric intensity is low enough, the uniform magnetic field will be in a prominent place, the assumed force is much bigger than the medial electric force. It will accelerate electrons and have enough time to thrust them escaping along the magnetic field line, and then decrease electrons' number reaching the collector and cut down the collected current in Figure 6. Although it will also increase the electrons' energy to reach the collector by the interparticle collision.

The results obtained indicate that the impact of magnetic field on the collected current in the F-H experiment is different even contrary distinctly, when the accelerate electric field and applied magnetic field are situated on different intensity stages.

\psection{CONCLUSIONS}

In the F-H experiment, some complicated impacts of applied magnetic field on the collected current have been observed. Increasing the applied magnetic field will lift up the electron kinetic energy, and enlarge the number of escaping electrons. Consequently increasing the applied magnetic field has a quite strange and complex influence on the collected current.

The impact of magnetic field variation on the collected current in the F-H test is equal to that of temperature variety. On the one hand, in case the accelerating electric intensity occupies a dominant position, enlarging magnetic flux density is equivalent to increasing of temperature. On the other hand, when the accelerating electric intensity becomes weak enough and applied magnetic flux density is in a prominent place, enhancing magnetic flux density should be identical with temperature decreasing.

It should be noted that the study for impacts of applied magnetic field on F-H test examined only some simple cases in the room temperature. Despite its preliminary characteristics, this study can clearly indicate that the applied magnetic flux density has an influence on the inelastic collisions of electrons with argon atoms. For the future studies, the research will concentrate on only more inferences about the impact of applied magnetic field on collected currents.

\ack
This project was supported partially by the National Natural Science Foundation of China under grant number 60677039.

\end{paper}
%--------------------------


\begin{thebibliography}{99}

%1
\bibitem{franck}
      Franck,~J. and Hertz,~G.,
      ``Uber Zusammenstobe zwischen Elektronen und Molek¨¹len des Quecksilberdampfes und die Ionisierungsspannung desselben",
      {\it Verh. Dtsch. Phys. Ges.\/},
      Vol.~16, 456--457, 1914.
      [English translation: World of the atom, edited by H. A. Boorse and L. Motz (Basic Books, New York, 1966), Vol.I, pp.770-778.]

%2
\bibitem{davis}
      Davis,~B. and Goucher,~F.~S.,
      ``Ionization and Excitation of Radiation by Electron Impact in Mercury Vapor and Hydrogen",
      {\it Physical Review\/},
      Vol.~10, No.~2, 101--115, 1917.

%3
\bibitem{liu}
      Liu,~F.-H.,
      ``Franck-Hertz experiment with higher excitatin level measurements",
      {\it American Journal of Physics\/},
      Vol.~55, No.~4, 366--369, 1987.

%4
\bibitem{robson}
      Robson,~R.~E., Li,~B. and White,~R.~D.,
      ``Spatially periodic structures in electron swarms and the Franck-Hertz experiment",
      {\it Journal of Physics B: Atomic, Molecular and Optical Physics\/},
      Vol.~33, No.~3, 507--520, 2000.

%5
\bibitem{hamilton}
      Hamilton,~W.~R.,
      {\it Elements of Quaternions\/},
      Longmans, Green \& Co., London, 1866.

%6
\bibitem{maxwell}
      Maxwell,~J.~C.,
      {\it A Treatise on Electricity and Magnetism\/},
      Dover Publications Inc., New York, 1954.

%7
\bibitem{weng}
      Weng,~Z.-H.,
      ``Electromagnetic forces on charged particles",
      {\it PIERS Proceedings\/}, 361--363,
      Moscow, Russia, August 18-21, 2009.


\end{thebibliography}
\end{document}